\documentclass[9pt,twocolumn,twoside]{pnas-new}

\templatetype{pnasresearcharticle} 

\definecolor{blue}{rgb}{0,0,1}

\definecolor{red}{rgb}{1,0,0}

\usepackage{subfigure}
\usepackage{adjustbox}

\title{Rate-induced tipping in complex high-dimensional ecological networks}

\author[a]{Shirin Panahi}
\author[b]{Younghae Do}
\author[c]{Alan Hastings}
\author[a,d,1]{Ying-Cheng Lai}

\affil[a]{School of Electrical, Computer and Energy Engineering, Arizona State University, Tempe, Arizona 85287, USA}
\affil[b]{Department of Mathematics, Nonlinear Dynamics $\&$ Mathematical Application Center, Kyungpook National University, Daegu 41566, Republic of Korea}
\affil[c]{Department of Environmental Science and Policy, University of California, One Shields Avenue, Davis, CA 95616, USA and Santa Fe Institute, 1399 Hyde Park Road, Santa Fe, NM 87501, USA}
\affil[d]{Department of Physics, Arizona State University, Tempe, Arizona 85287, USA}

\leadauthor{Panahi}

\significancestatement{
	Human activities are having increasingly negative impacts on natural systems and it is of interest to understand how the ``pace'' of parameter change may lead to catastrophic consequences. This work studies the phenomenon of rate-induced tipping (R-tipping) in high-dimensional ecological networks, where the rate of parameter change can cause the system to undergo a tipping point from healthy survival to extinction. A quantitative scaling law between the probability of R-tipping and the rate was uncovered, with a striking and devastating consequence: in order to reduce the probability, parameter changes must be slowed down to such an extent that the rate practically reaches zero. This may pose an extremely significant challenge in our efforts to protect and preserve the natural environment.}

\authorcontributions{YCL and AH conceived the idea. SP performed computations. All analyzed data. YCL and SP wrote the paper. YCL and AH edited the paper.}
\authordeclaration{The authors declare no competing interests.}
\correspondingauthor{\textsuperscript{1}To whom correspondence should be addressed. E-mail: Ying-Cheng.Lai@asu.edu}


\begin{abstract}

In an ecosystem, environmental changes as a result of natural and human processes can cause some key parameters of the system to change with time. Depending on how fast such a parameter changes, a tipping point can occur. Existing works on rate-induced tipping, or R-tipping, offered a theoretical way to study this phenomenon but from a local dynamical point of view, revealing, e.g., the existence of a critical rate for some specific initial condition above which a tipping point will occur. As ecosystems are subject to constant disturbances and can drift away from their equilibrium point, it is necessary to study R-tipping from a global perspective in terms of the initial conditions in the entire relevant phase space region. In particular, we introduce the notion of the probability of R-tipping defined for initial conditions taken from the whole relevant phase space. Using a number of real-world, complex mutualistic networks as a paradigm, we discover a scaling law between this probability and the rate of parameter change and provide a geometric theory to explain the law. The real-world implication is that even a slow parameter change can lead to a system collapse with catastrophic consequences. In fact, to mitigate the environmental changes by merely slowing down the parameter drift may not always be effective: only when the rate of parameter change is reduced to practically zero would the tipping be avoided. Our global dynamics approach offers a more complete and physically meaningful way to understand the important phenomenon of R-tipping.  

\end{abstract}

\dates{This manuscript was compiled on \today}
\doi{\url{www.pnas.org/cgi/doi/10.1073/pnas.XXXXXXXXXX}}

\begin{document}

\maketitle
\thispagestyle{firststyle}
\ifthenelse{\boolean{shortarticle}}{\ifthenelse{\boolean{singlecolumn}}{\abscontentformatted}{\abscontent}}{}

\dropcap{I}n complex dynamical systems, the phenomenon of tipping point, 
characterized by an abrupt transition from one type of behavior (typically 
normal, healthy, or survival) to another type (e.g., catastrophic), has 
received growing attention~\cite{Scheffer:2004,Schefferetal:2009,Scheffer:2010,WH:2010,DJ:2010,CLLLA:2012,BH:2012,DVKG:2012,ashwin2012tipping,LLDvanNS:2012,Barnoskyetal:2012,BH:2013,TC:2014,LNSB:2014,LCJL:2015,GTZB:2015,JHSLGHL:2018,YLTLZCX:2018,JHL:2019,scheffer2020critical,MJGL:2020,MLG:2020,MG:2021,MLG:2022,o2020tipping}.
A tipping point is often of significant concern because it is a point of 
``no return'' in the parameter space, manifested by the collapse of the system
as a parameter passes through a critical value. In ecological systems, sudden 
extinction of species on a large scale can be the result of a tipping 
point~\cite{Scheffer:2004,Schefferetal:2009,Scheffer:2010,WH:2010}. 
Tipping points can arise in diverse contexts such as the outbreak of 
epidemics~\cite{trefois2015critical}, global climate 
changes~\cite{albrich2020climate}, and the sudden transition from normal to 
depressed mood in bipolar patients~\cite{bayani2017critical}. In nonlinear 
dynamics, a common type of bifurcation responsible for a tipping point is 
saddle-node bifurcations (forward or backward). Consider the situation where, 
in the parameter regime of interest, there are two coexisting stable steady 
states: a ``high'' state corresponding to normal or ``survival'' functioning 
and a ``low'' or ``extinction'' state, each with its own basin of attractor.
Suppose external factors such as climate change cause a bifurcation parameter 
of the system to increase. A tipping point is a backward saddle-node 
bifurcation, after which the ``survival'' fixed point disappears, leaving 
the ``extinction'' state as the only destination of the system, where the 
original basin of the ``survival'' state is absorbed into the basin of the
``extinction'' state. 

In real-world dynamical systems, parameters are not stationary but constantly 
drift in time. A known example is the slow increase in the average global 
temperature with time due to human activities. In ecological systems, some key
parameters such as the carrying capacity or the species decay rate can change 
with time, and there is a global tendency for such parameter changes to 
``speed up.'' In fact, the rate of environmental change is an important driver
across different scales in ecology~\cite{synodinos2022rate}. The behavior of 
the parameter variations with time introduces another ``meta'' parameter into 
the dynamical process: the rate of change of the parameters. About ten years 
ago, it was found that the rate can induce a tipping point - the phenomenon of
rate-induced tipping or R-tipping~\cite{ashwin2012tipping}, which is relevant 
to fields such as climate science~\cite{morris2002responses, ritchie2023rate}, 
neuroscience~\cite{wieczorek2011excitability,mitry2013excitable}, vibration 
engineering~\cite{alexander2011exploring}, and even competitive 
economy~\cite{hsu2015tendency}. 
{Existing studies of R-tipping were for low-dimensional dynamical systems and the 
analysis approaches were ``local'' in the sense that they focused on the 
behaviors of some specific initial condition and trajectory, addressing issues 
such as the critical rate for tipping~\cite{ashwin2017parameter,vanselow2019very}.
In particular, with respect to a specific initial condition, R-tipping in these 
studies was defined as an abrupt change in the behavior of the system (or a 
critical transition) taking place at a specific rate of change of a bifurcation 
parameter~\cite{ashwin2012tipping}.}

The state that an ecological system is in depends on a combination of deterministic dynamics, small scale stochastic influences (e.g., demographic stochasticity~\cite{lande2003}), and large stochastic disturbances such as drought or other significant climatic event~\cite{hansen2019}. So when considering future dynamics of ecological systems it makes sense to consider systems that may be far from equilibrium, but still within the basin of attraction of an equilibrium, rather than starting at the equilibrium.  High-dimensional ecological systems are particularly likely to be found far from equilibrium. In fact, it has been suggested that it can be common for ecosystems to be in some transient state~\cite{HACFGLMPSZ:2018,MACFGHLPSZ:2020}.

In this paper, we focus on high-dimensional, empirical ecological networks and
investigate R-tipping with two key time-varying parameters by presenting a
computationally feasible, ``global'' approach to exploring the entire relevant
phase space region with analytic insights.
We focus on a representative class of such systems: 
mutualistic networks~\cite{BJMO:2003,GJT:2011,NJB:2013,LNSB:2014,RSB:2014,DB:2014,GPJBT:2017,JHSLGHL:2018,MJGL:2020,JHL:2019,OSH:book} that are fundamental to 
ecological systems, which are formed through mutually beneficial interactions 
between two groups of species. In a mutualistic network, a species in one 
group derives some form of benefit from the relationship with some species in 
the other group. Examples include the corals and the single-celled 
zooxanthellae that form the large-scale coral reefs, and the various networks of 
pollinators and plants in different geographical regions of the world. These 
networks influence biodiversity, ecosystem stability, nutrient cycling, community 
resilience, and evolutionary dynamics~\cite{bastolla2009architecture}, and they
are a key aspect of ecosystem functioning with implications for conservation and 
ecosystem management. Understanding the ecological significance of mutualistic 
networks is crucial for unraveling the complexities of ecological communities and 
for implementing effective strategies to safeguard biodiversity and ecosystem 
health~\cite{bascompte2007plant}. Mathematically, because of the typically large 
number of species involved in the mutualistic interactions, the underlying 
networked systems are high-dimensional nonlinear dynamical systems.

We first ask if R-tipping can arise in such high-dimensional systems through 
simulating a number of empirical pollinator-plant mutualistic networks (see 
Tab.~\ref{Tab:Net}) and obtain an affirmative answer.
{Our computations reveal that the critical rate above which a tipping point occurs
depends strongly on the initial condition. Rather than studying the critical rate
for any specific initial condition, we go beyond the existing local analysis
approaches by investigating the {\em probability of R-tipping for a large ensemble
of initial conditions taken from the whole relevant phase space} and asking how
this probability, denoted as $\Phi (r)$, depends on the rate $r$ of parameter 
change.}
We discover a scaling law between $\Phi (r)$ and $r$: as the rate increases, the 
probability first increases rapidly, then slowly, and finally saturates. 
Using a universal two-dimensional (2D) effective model that was validated to be 
particularly suitable for predicting and analyzing tipping points in 
high-dimensional mutualistic networks~\cite{JHSLGHL:2018}, we analytically 
derive the scaling law. Specifically, let $\kappa (t)$ be the parameter that 
changes with time linearly at the rate $r$ in a finite range, denoted as 
$[\kappa_{\rm min},\kappa_{\rm max}]$. The scaling law is given by
\begin{align} \label{eq:scaling}
\Phi (r) \sim \exp{ [- C (\kappa_{\rm max} - \kappa_{\rm min})/r]},
\end{align}
where $C > 0$ is a constant. Our theoretical analysis indicates that, for 
2D systems, $C$ is nothing but the maximum possible unstable 
eigenvalue of the mediating unstable fixed point on the boundary separating 
the basins of the survival and extinction attractors when the parameter 
$\kappa$ varies in the range $[\kappa_{\rm min},\kappa_{\rm max}]$. However, 
for high-dimensional systems, such correspondence does not hold, but $C$ can 
be determined through a numerical fitting. 

The scaling law (\ref{eq:scaling}) has the following features. First, the 
probability $\Phi (r)$ is an increasing function of $r$ for $r > 0$ 
[$\Phi'(r) > 0$]. Second, $\Phi'(r)$ is a decreasing function of $r$, i.e., 
the increase of $\Phi (r)$ with $r$ slows down with $r$ and the rate of 
increase becomes zero for $r\rightarrow\infty$. Third, the rate of increase 
in $\Phi (r)$ with $r$ is the maximum for $r\gtrsim 0$, and $\Phi (r)$ becomes 
approximately constant for  
$r > r^* \sim \sqrt{(\kappa_{\rm max} - \kappa_{\rm min})}$. This third feature
has a striking implication because the probability of R-tipping 
will grow dramatically as soon as the rate of parameter change increases from 
zero. The real-world implication is alarming because it means that even a slow 
parameter change can lead to a system collapse with catastrophic consequences. 
To control or mitigate the environmental changes by merely slowing down the 
parameter drift may not always be effective: only when the rate of parameter 
change is reduced to practically zero would the tipping be avoided!

Alternatively, the scaling law (\ref{eq:scaling}) can be expressed as the 
following explicit formula:
\begin{align} \label{eq:scaling_formula}
\Phi (r) = B \exp{ [-C (\kappa_{\rm max} - \kappa_{\rm min})/r]}
\end{align}
with an additional positive constant $B$. For 2D systems, $B$ is
the difference between the basin areas of the extinction state for 
$\kappa = \kappa_{\rm max}$ and $\kappa = \kappa_{\rm min}$. For 
high-dimensional systems, $B$ can be determined numerically.

{While a comprehensive understanding of the entire parameter space as well as the 
rates and directions of change associated with these parameters are worth 
investigating, the picture that motivated our work was the potential impacts of 
ongoing climate change on ecological systems. Under climate change, various 
parameters can vary in different directions, and some changes might offset the 
effects of others. Our focus is on the scenarios where the changes in parameters 
align with the observed detrimental environmental impacts. For example, consistent 
with environmental deterioration, in a mutualistic network the species decay 
rate can increase, and/or the mutualistic interaction strength can decrease. A 
simultaneous increase in the species decay rate and mutualistic strength does 
not seem physically reasonable in this context. We thus study scenarios of multiple 
parameter variations that align with the realistic climate change impacts by carrying 
out computations with a systematic analysis of the tipping-point transitions in the 
two-dimensional parameter plane of the species decay rate and the mutualistic 
interaction strength.}

\section*{Results} 

{We consider the empirical mutualistic pollinator–plant networks from the 
Web of Life database (\href{https://www.web-of-life.es/}{www.Web-of-Life.es}).
The needs to search for the parameter regions exhibiting R-tipping and to simulate 
a large number of initial conditions in the phase space as required by our global 
analysis approach, as well as the high dimensionality of the empirical mutualistic 
networks demand extremely intensive computation~\cite{menck2013basin}. 
To make the computation feasible, 
we select ten networks to represent a diverse range of mutualistic 
interactions from different regions of the world and highlight the generality of our 
approach to R-tipping and the scaling law.} 
The basic information about these networks such as the name of the networks, the 
countries where the empirical data were collected, and the number of species in each 
network, is listed in Tab.~\ref{Tab:Net}. The dynamics of a mutualistic network of 
$N_A$ pollinator and $N_P$ plant species, taking into account the generic 
plant-pollinator interactions~\cite{BJMO:2003}, can be described by a 
total of $N = N_A + N_P$  nonlinear differential equations of the Holling 
type~\cite{holling1973resilience} in terms of the species abundances as 
\begin{subequations}
\begin{align} \label{eq:dPdt}
    \dot P_i= & P_i \Big ( \alpha_i^P - \sum_{l=1}^{N_p} \beta_{il}^{P} P_l + \frac{\sum_{j=1}^{N_A} \gamma_{ij}^P A_j}{1+h\sum_{j=1}^{N_A} \gamma_{ij}^P A_j}  \Big), \\ \label{eq:dAdt}
    \dot A_j= & A_j\Big ( \alpha_j^A -\kappa_j- \sum_{l=1}^{N_A} \beta_{jl}^{A} A_l + \frac{\sum_{i=1}^{N_P} \gamma_{ji}^A P_i}{1+h\sum_{i=1}^{N_P} \gamma_{ji}^A P_i}  \Big),
\end{align}
\end{subequations}
where $P_i$ and $A_j$ are the abundances of the $i^{th}$ and $j^{th}$ plant and
pollinator species, respectively, $i=1,\ldots,N_P$, $j=1,\ldots,N_A$, $\kappa$ 
is the pollinator decay rate, $\alpha^{P(A)}$ is the intrinsic growth rate 
in the absence of intraspecific competition and any mutualistic effect, $h$ is 
the half-saturation constant. Intraspecific and interspecific competition of 
the plants (pollinators) is characterized by the parameters $\beta_{ii}^{P}$ 
($\beta_{jj}^{A}$) and $\beta_{il}^{P}$ ($\beta_{jl}^{A}$), respectively. 
The mutualistic interactions as characterized by the parameter $\gamma_{ij}^P$
can be written as $\gamma_{ij}^P=\xi_{ij}\gamma_0/K_i^{\tau}$, where 
$\xi_{ij}=1$ if there is a mutualistic interaction between the $i^{th}$ plant 
and $j^{th}$ pollinator (zero otherwise), $\gamma_0$ is a general interaction 
parameter, $K_i$ is the degree of the $i^{th}$ plant species in the network, 
and $\tau$ determines the strength of the trade-off between the interaction 
strength and the number of interactions. (The expression for $\gamma_{ji}^A$
is similar.)

\begin{table}[H]
\caption{Empirical mutualistic networks studied in this work and their corresponding range of parameter change and fitting constants in Eq.~(\ref{eq:scaling})}
\centering
\begin{tabular}{c|c|c|c|c|c|c}
 Network & Country &  $N_P$ &  $N_A$  &  $\kappa$-interval  &  $B$  &  $C$ \\
	\hline
        \hline
        {M\_PL\_008} & {Canary Islands} & {$11$} & {$38$} & ${[0.90,0.93]}$ & {$0.36$} & {$0.19$} \\\hline
        {M\_PL\_013} & {South Africa} & {$9$} & {$56$} & ${[0.70,0.99]}$ & {$0.07$} & {$0.11$} \\\hline
        {M\_PL\_022} & {Argentina} & {$21$} & {$45$} & ${[0.75,0.93]}$ & {$0.88$} & {$0.12$} \\\hline
	M\_PL\_023 & Argentina & $23$ & $72$ & $[0.87,0.96]$ & $0.59$ & $0.09$ \\\hline
	M\_PL\_027 & New Zealand & $18$ & $60$ & $[0.90,0.95]$ & $0.49$ & $0.09$ \\\hline
        M\_PL\_032 & USA & $7$ & $33$ & $[0.82,0.99]$ & $0.55$ & $0.19$ \\\hline  
        M\_PL\_036 & A\c{c}ores & $10$ & $12$ & $[0.74,0.88]$ & $0.39$ & $0.12$ \\\hline 
        M\_PL\_037 & Denmark & $10$ & $40$ & $[0.87,0.93]$ & $0.23$ & $0.15$ \\\hline
        M\_PL\_038 & Denmark & $8$ & $42$ & $[0.85,0.95]$ & $0.30$ & $0.15$ \\\hline
        M\_PL\_045 & Greenland & $17$ & $26$ & $[0.96,0.98]$ & $0.09$ & $0.05$
\end{tabular} \label{Tab:Net}
\end{table}

The computational setting of our study is as follows. In the network system
described by Eqs.~\ref{eq:dPdt} and \ref{eq:dAdt}, 
the number of the equations determines the
phase-space dimension of the underlying nonlinear dynamical system. There are
a large number of basic parameters in the model, such as $\gamma$ that quantifies
the strength of the mutualistic interactions and $\kappa$ characterizing the
species decay rate. In the context of R-tipping, while all the parameters should
be time-varying in principle, to make our study computationally feasible, we
assume that the two key parameters ($\kappa$ and $\gamma$) are time-dependent
while keeping the other parameters fixed. Since the defining characteristic of
a system exhibiting a tipping point is the coexistence of two stable steady
states: survival and extinction, we focus on the range of parameter variations
in which the network system under study exhibits the two stable equilibria.
When presenting our results (Fig.~\ref{fig:prnet} below 
and Figs.~S4-S14 in Supplementary
Information), in each case the parameter variation is along a specific direction
in the two-dimensional parameter plane: $\kappa$ and $\gamma$ with the goal to
understand how changes in these two parameters impact the R-tipping probability.

To introduce the rate change of a parameter, we consider the scenario where 
negative environmental impacts cause the species decay rate to increase 
linearly with time at the rate $r$ from a minimal value $\kappa_{\rm min}$ to
a maximal value $\kappa_{\rm max}$: 
\begin{align} \label{eq:rate}
\dot \kappa_j = \begin{cases}
r \quad {\rm if} \quad \kappa_{\rm min} < \kappa_j < \kappa_{\rm max}\\
0 \quad\quad {\rm otherwise.}
\end{cases}
\end{align}
(For mutualistic networks, another relevant parameter that is vulnerable to 
environmental change is the mutualistic interaction strength. The pertinent 
results are presented in Supplementary Information.)
To calculate the probability of R-tipping, $\Phi (r)$, we set $r = 0$ so that
$\kappa = \kappa_{\rm min}$, solve Eqs.~(\ref{eq:dPdt}) and (\ref{eq:dAdt})
numerically for a large number of random initial conditions chosen uniformly 
from the whole high-dimensional phase space, and determine $10^5$ initial 
conditions that approach the high stable steady state in which no species 
becomes extinct. We then increase the rate $r$ from zero. For each fixed value 
of $r$, we calculate, for each of the selected $10^5$ initial conditions, 
whether or not the final state is the high stable state. If yes, then there is 
no R-tipping for the particular initial condition. However, if the final state 
becomes the extinction state, R-tipping has occurred for this value of $r$. The 
probability $\Phi (r)$ can be approximated by the fraction of the number of 
initial conditions leading to R-tipping out of the $10^5$ initial conditions.  

\begin{figure*}[ht!]
\centering
\includegraphics[width=\linewidth]{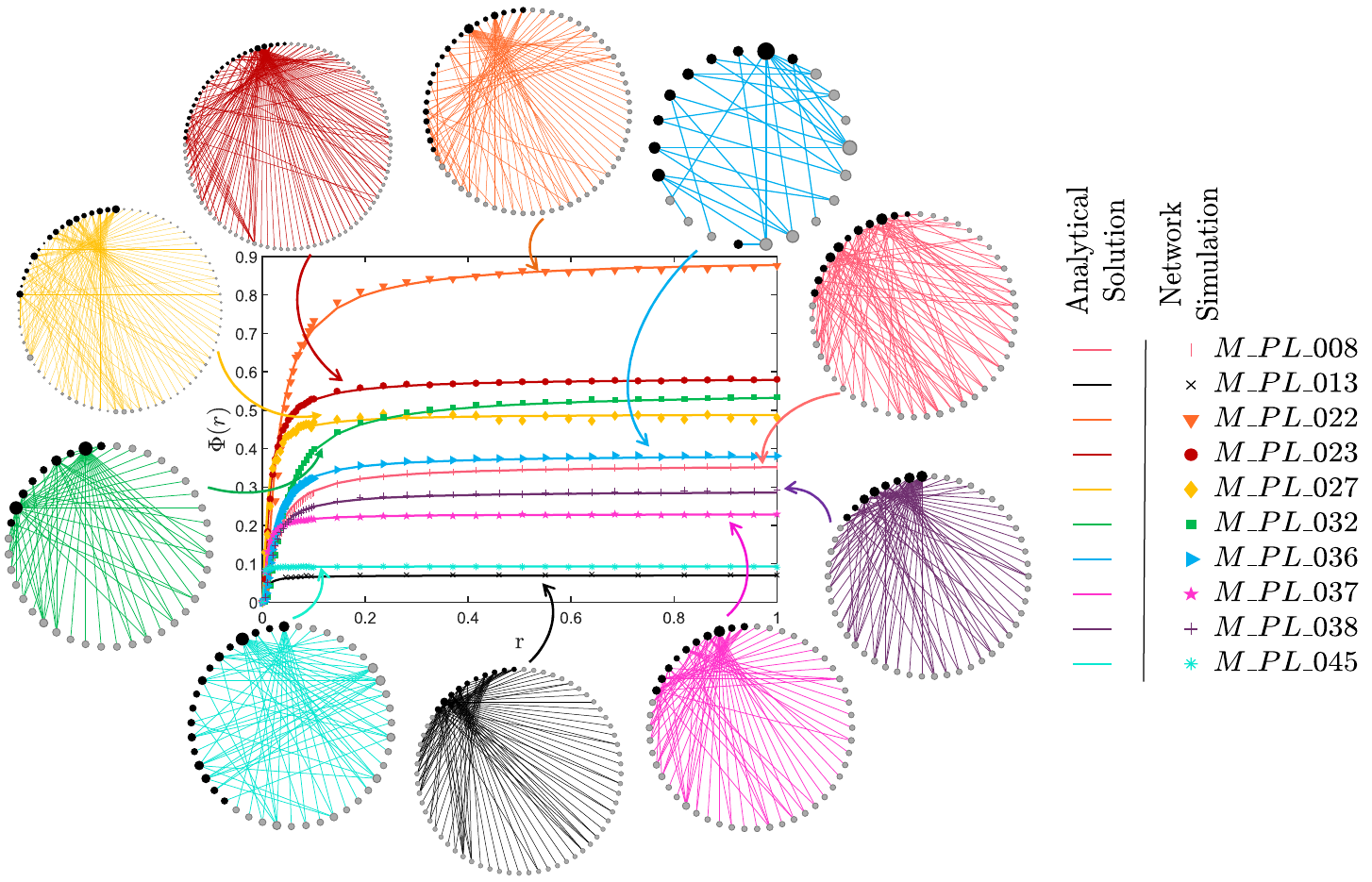}
\caption{ 
Probability $\Phi (r)$ of R-tipping versus the time rate $r$ of parameter change 
for ten real-world mutualistic networks, together with the corresponding bipartite 
structure for each network. The dots are the probability calculated by simulating 
Eqs.~(\ref{eq:dPdt}), (\ref{eq:dAdt}), and (\ref{eq:rate}), with an ensemble of 
random initial conditions. The solid curves are the analytic results from 
Eq.~(\ref{eq:scaling}), with the two fitting parameters $B$ and $C$ given in 
Tab.~\ref{Tab:Net}. Other parameter values are $\alpha=0.3$, $\beta=1$, $h=0.4$, 
$\gamma^P=1.93$, and $\gamma^A=1.77$.}
\label{fig:prnet}
\end{figure*}

Figure~\ref{fig:prnet} shows $\Phi (r)$ versus $r$ for the ten empirical mutualistic 
networks specified in Tab.~1, together with the bipartite network structure, which
are distinguished by different colors. For each network, the interval
$[\kappa_{\rm min},\kappa_{\rm max}]$ of the bifurcation parameter (listed in
the fifth column of Tab.~\ref{Tab:Net}) 
is chosen such that the dynamical network for static
parameter values exhibits two stable steady states and a tipping point
in this interval, as determined
by a computational bifurcation analysis of the species abundances versus $\kappa$.
Despite the differences in the topology and the specific parameter values among
the empirical networks, it is remarkable that the probability of R-tipping
exhibits a characteristic behavior common among all the networks: as the
time-varying rate of the bifurcation parameter increases from zero, the
probability increases rapidly and then saturates at an approximately finite
constant value, as quantified by the analytic scaling law (\ref{eq:scaling}). 
The final or
asymptotic value of the R-tipping probability attained in the regime of large
time rate of change of the bifurcation parameter depends on the specifics of
the underlying mutualistic network in terms of its topological structure and
basic system parameters. For example, the total number of species, or the
phase-space dimension of the underlying dynamical system, and the relative
numbers of the pollinator and plant species vary dramatically across the ten
networks, as indicated by the ten surrounding network-structure diagrams in
Fig.~\ref{fig:prnet}. 
The structural and parametric disparities among the networks lead to
different relative basin volumes of the survival and extinction stable steady
states, giving rise to distinct asymptotic probabilities of R-tipping.

The ecological interpretation of the observed different asymptotic values of the
R-tipping probability is, as follows. Previous studies revealed the complex
interplay between the structural properties of the network and environmental
factors in shaping extinction probabilities~\cite{schleuning2016ecological,vanbergen2017network,lewinsohn2006structure,mcp057,aslan2013mutualism,sheykhali2020robustness,pires2020indirect,young2021reconstruction}. For example, it was
discovered~\cite{schleuning2016ecological} that the plant species tend to play a more significant role in
the health of the network system as compared to pollinators~\cite{schleuning2016ecological} in the sense
that plant extinction due to climate change is more likely to trigger pollinator
coextinction than the other way around. In another example, distinct topological
features were found to be associated with the networks with a higher probability
of extinction~\cite{vanbergen2017network,lewinsohn2006structure}. It was also found that robust plant-pollinator
mutualistic networks tend to exhibit a combination of compartmentalized and
nested patterns~\cite{lewinsohn2006structure}. More generally, mutualistic networks in the real world
are quite diverse in their structure and parameters, each possessing one or two
or all the features including lower interaction density, heightened
specialization, fewer pollinator visitors per plant species, lower nestedness, and
lower modularity, etc.~\cite{vanbergen2017network,lewinsohn2006structure,mcp057,aslan2013mutualism,sheykhali2020robustness,pires2020indirect,young2021reconstruction}. It is thus reasonable to hypothesize that the
asymptotic value of the R-tipping probability can be attributed to the
sensitivity of the underlying network to environmental changes. For example,
a higher (lower) saturation value of the extinction probability can be associated
with the networks with a large (small) number of plants and lower (higher)
nestedness, network $M\_PL\_22$ (network $M\_PL\_13$). In other examples,
network $M\_PL\_32$ can be categorized as ``vulnerable'' due to its low
modularity, despite having a small number of plants, and network $M\_PL\_45$ is
resilient against extinction due to its high nestedness and high modularity.

The remarkable phenomenon is that, in spite of these differences, the rapid initial 
increase in the R-tipping probability is shared by all ten networks!
That is when a parameter begins to change with time from zero, even slowly, the 
probability of R-tipping increases dramatically. The practical implication is that 
ecosystems with time-dependent parameter drift are highly susceptible to R-tipping. 
Parameter drifting, even at a slow pace, will be detrimental. This poses a daunting 
challenge to preserving ecosystems against negative environmental changes. In 
particular, according to conventional wisdom, ecosystems can be effectively 
protected by slowing down the environmental changes, but our results suggest that 
catastrophic tipping can occur with a finite probability unless the rate of 
the environmental changes is reduced to a near zero value.

\begin{figure}[h]
\centering
\includegraphics[width=\linewidth]{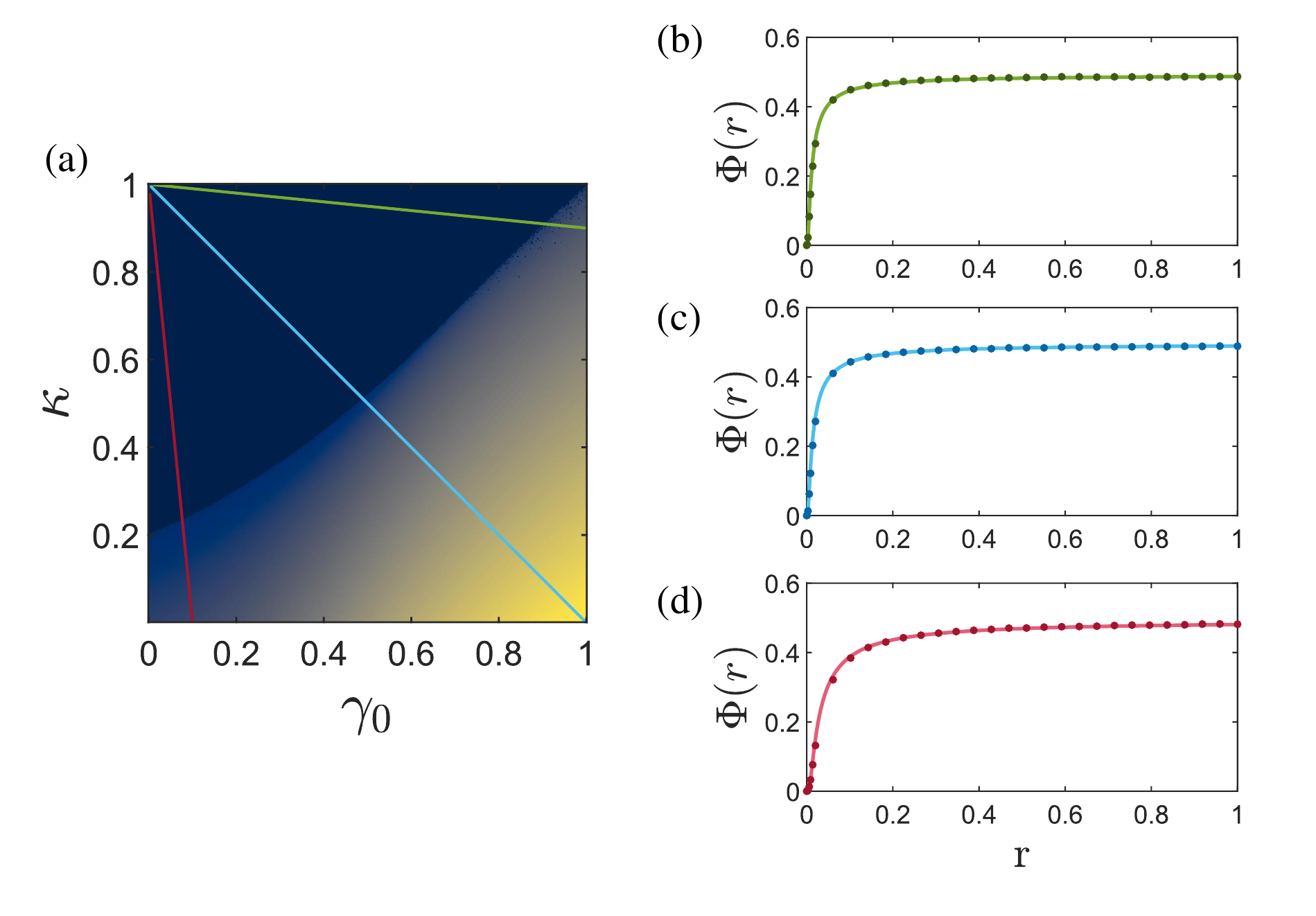}
\caption{ {Scaling law of probability of R-tipping in two-dimensional parameter space.
(a) Two-dimensional parameter plane of the species decay rate ($\kappa$) and the
mutualistic interaction strength ($\gamma_0$) for the network 
${\rm M}\_{\rm PL}\_032$. The
solid lines correspond to the three different scenarios of the parameter
$\mathcal{C}$ in calculating the probability of the R-tipping point from $10^7$
initial conditions for (b) $\mathcal{C}<1$, (c) $\mathcal{C}=1$, and
(d) $\mathcal{C}>1$. Other parameters are $t=0.5$, $\beta=1$, $\alpha=0.3$,
$\mu = 0$, and $h = 0.4$.}}
\label{fig:4d_32_all}
\end{figure}

{To further explore the effects of parameter changes, we study a scenario in which
negative environmental impacts cause the species decay rate ($\kappa$) and the
mutualistic interaction strength ($\gamma_0$) to linearly increase and decrease,
respectively. By considering the rate of changing of $\kappa$ ($\gamma_0$) as
$r_{\kappa}$ ($r_{\gamma_0}$), we set $r_{\kappa}=\mathcal{C} r_{\gamma_0}$, which
allows us to investigate three different scenarios by varying the parameter 
$\mathcal{C}$: $\mathcal{C}<1$, $\mathcal{C}=1$, and $\mathcal{C}>1$. 
Figure~\ref{fig:4d_32_all} shows a systematic analysis of the tipping-point 
transitions in the 2D parameter plane along with the probability of R-tipping for 
the three different scenarios for the network ${\rm M}\_{\rm PL}\_032$. (The 
pertinent results for the other nine networks are presented in Sec.~4 in 
Supporting Information.)}

To understand the behavior of the R-tipping probability in Fig.~\ref{fig:prnet},
we resort to the approach of dimension reduction~\cite{gao2016universal}. In 
particular, for tipping-point dynamics, a high-dimensional mutualistic network 
can be approximated by an effective model of two dynamical variables: the 
respective mean abundances of all pollinator and plant species. The effective
2D model can be written as~\cite{JHSLGHL:2018}
\begin{subequations}
	\begin{align} \label{eq:dPdt_2D}
\dot P &= P \Big (\alpha - \beta P + \frac{\gamma^P A}{1+h\gamma^P A}\Big), \\ \label{eq:dAdt_2D}
\dot A &= A\Big ( \alpha -\kappa- \beta A + \frac{\gamma^A P}{1+h\gamma^A P}  \Big),
\end{align}
\end{subequations}
where $P$ and $A$ are the average abundances of all the plants and pollinators,
respectively, $\alpha$ is the effective growth rate and parameter $\beta$ 
characterizes the combined effects of intraspecific and interspecific 
competition. The parameters $\gamma^P$ and $\gamma^A$ are the effective 
mutualistic interaction strengths that can be determined by the method of 
eigenvector weighting~\cite{JHSLGHL:2018} (Supplementary Note 1). 
Equations~(\ref{eq:dPdt_2D}) and (\ref{eq:dAdt_2D}) possess five possible 
equilibria: $\mathbf{f}_1 \equiv (0,0)^T$, 
$\mathbf{f}_2 \equiv (\alpha/\beta,0)^T$, 
$\mathbf{f}_3 \equiv (0,(\alpha - \kappa)/\beta)^T$, and
$\mathbf{f}_{4,5} \equiv (g_1,g_2)^T$, where $g_1$ and $g_2$ are two possible
equilibria that depend on the values of the parameters of the model and can be 
calculated by setting zero the factor in the parentheses of 
Eqs.~(\ref{eq:dPdt_2D}) and (\ref{eq:dAdt_2D}).  

\begin{figure}[ht!]
\centering
\includegraphics[width=\linewidth]{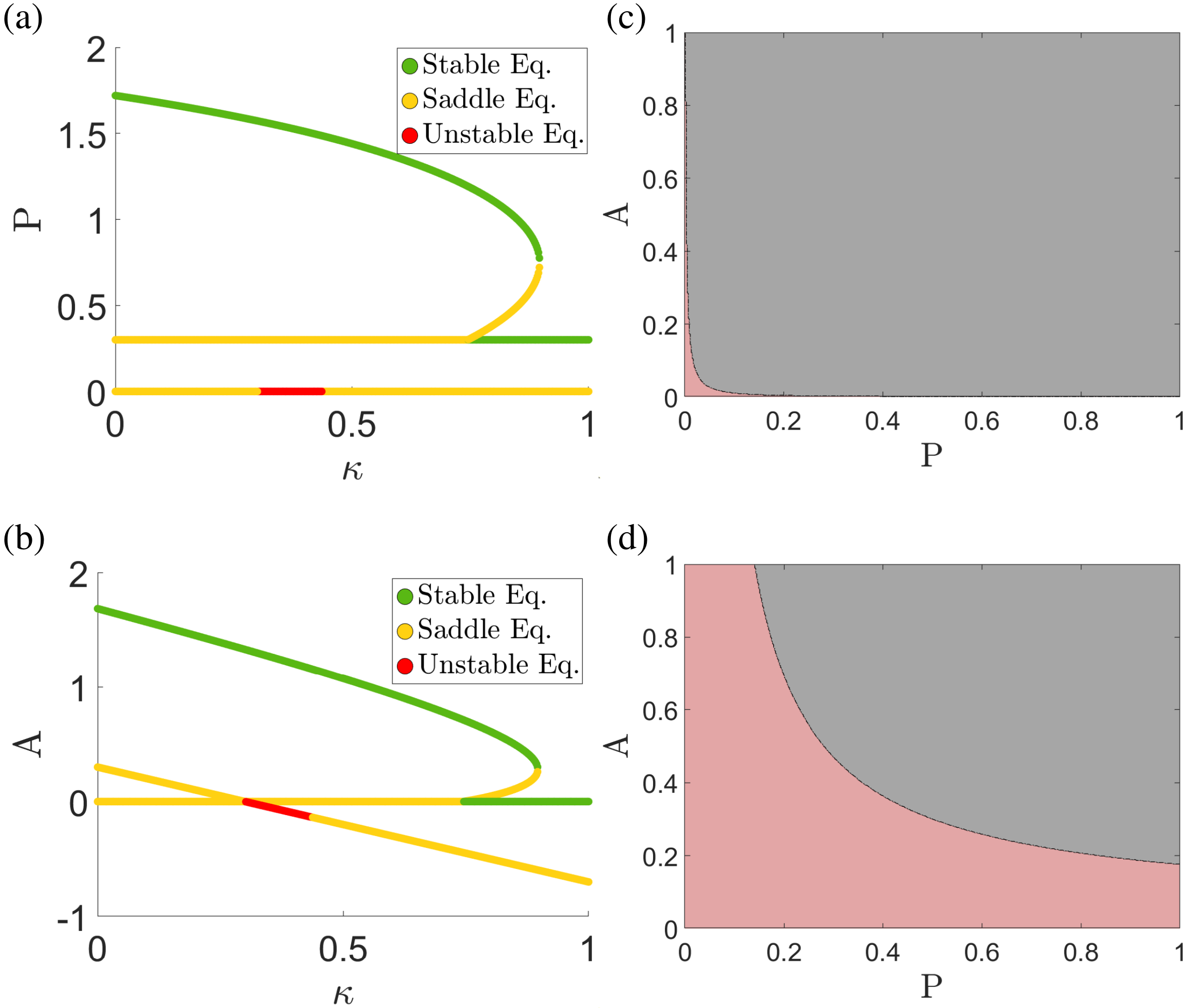}
\caption{Dynamics of the 2D reduced model Eqs.~(\ref{eq:dPdt_2D}) and 
(\ref{eq:dAdt_2D}). (a,b) A tipping point occurs as the species decay rate 
$\kappa$ increases towards one, for the mean plant and pollinator abundances, 
respectively. The green and yellow curves correspond to the stable and unstable
steady states, respectively, where the yellow curve with a red segment is
the unstable fixed point $\mathbf{f}_5$. The region of bistability is indicated
by the blue dot-dashed ellipses. (c,d) Examples of basins of attraction of the 
2D model whose parameters are determined as the corresponding averages from the 
empirical network ${\rm M}\_{\rm PL}\_036$ in Tab.~\ref{Tab:Net} for 
$\kappa_{min}=0.74$ and $\kappa_{max}=0.88$, respectively. Other parameters are 
$\alpha=0.3$, $\beta=1$, $h=0.4$, $\gamma^P=1.93$, and $\gamma^A=1.77$. The pink 
and gray regions correspond to the basins of the stable steady states $f_2$ 
(extinction) and $f_4$ (survival), respectively.}
\label{fig:equi}
\end{figure}

The first equilibrium $\mathbf{f}_1$, an extinction state, is at the origin 
$(P^*,\, A^*)=(0,0)$ and is unstable. The second equilibrium $\mathbf{f}_2$ 
is located at the $(P^*,\, A^*)=(\alpha/\beta,0)$ and it can be stable or 
unstable depending on the parameters $\kappa$. The locations of the remaining 
equilibria depend on the value of $\kappa$. In particular, the third 
equilibrium $\mathbf{f}_3$ can be unstable or nonexistent and the fourth 
equilibrium $\mathbf{f}_4$ is stable and coexists with stable equilibrium 
$\mathbf{f}_2$ in some interval of $\kappa$. The fifth equilibrium 
$\mathbf{f}_5$ is an unstable saddle fixed point in some relevant interval of 
$\kappa$. Figures~\ref{fig:equi}(a) and \ref{fig:equi}(b) exemplify the 
behaviors of the equilibria as $\kappa$ increases from zero to one for 
$\alpha=0.3$, $\beta=1$, $h=0.4$, $\gamma^P=1.93$, and $\gamma^A=1.77$,
for the average plant and pollinator abundances, respectively. In each panel,
the upper green curve is the survival fixed point $\mathbf{f}_4$, while the 
lower horizontal green line corresponds to the extinction state $\mathbf{f}_2$. 
The blue dot-dashed ellipse indicates the interval of $\kappa$ in which two 
stable equilibria coexist (bistability), whose right edge marks a tipping 
point. 

\begin{figure}[ht!]
\centering
\includegraphics[width=\linewidth]{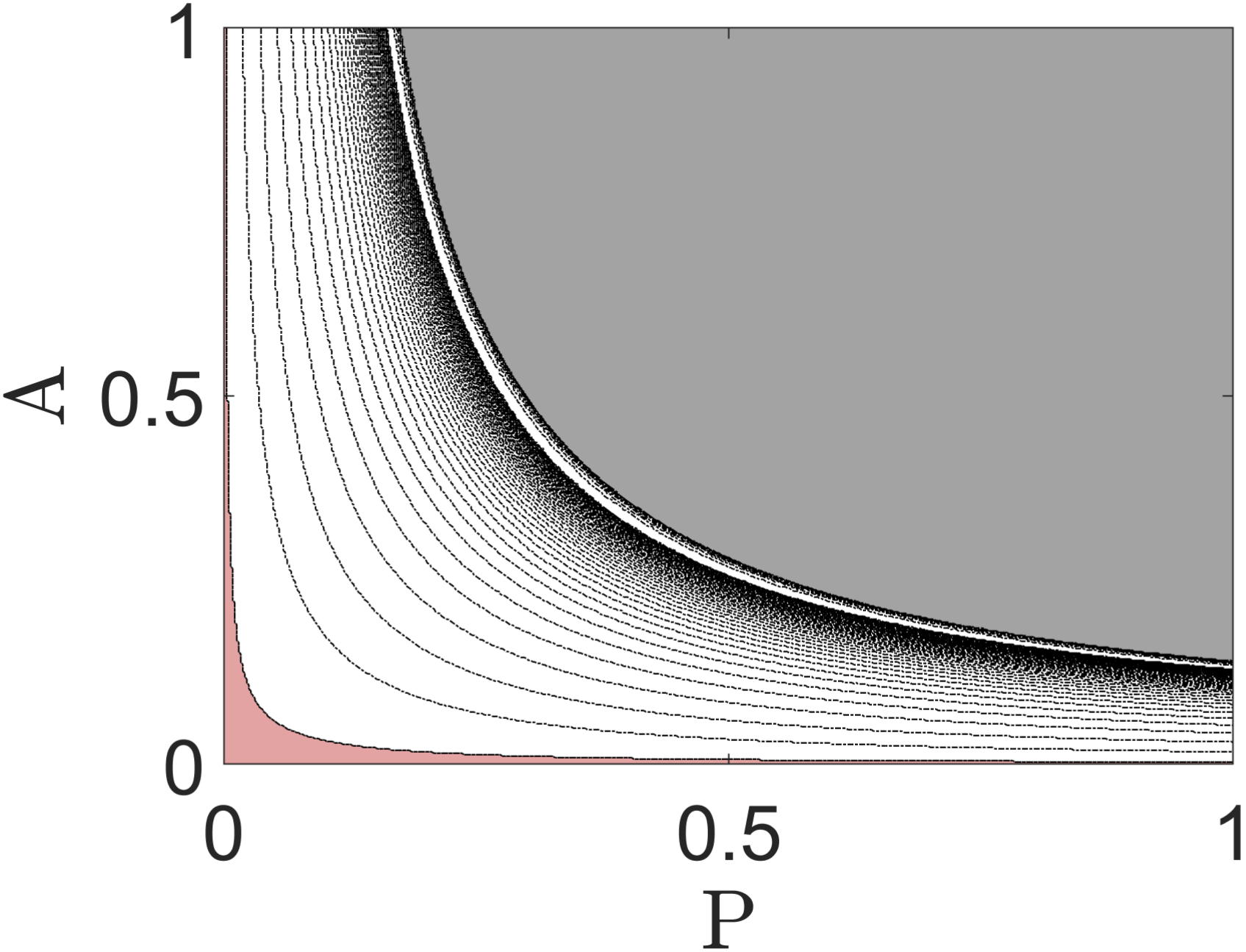}
\caption{Basin boundary between the extinction and survival fixed-point 
attractors for different rate of parameter change. Shown are a series of basin 
boundaries for $r \in [0\, 1]$. As $r$ increases from zero, the boundary moves 
in the direction of large species abundances. The basin boundaries are 
calculated from the 2D model of the empirical network ${\rm M}\_{\rm PL}\_036$ in
Tab.~\ref{Tab:Net}. Other parameters are the same as those in 
Figs.~\ref{fig:equi}(c) and \ref{fig:equi}(d).}
\label{fig:bb}
\end{figure}

\begin{figure*}[ht!]
\centering
\includegraphics[width=0.8\linewidth]{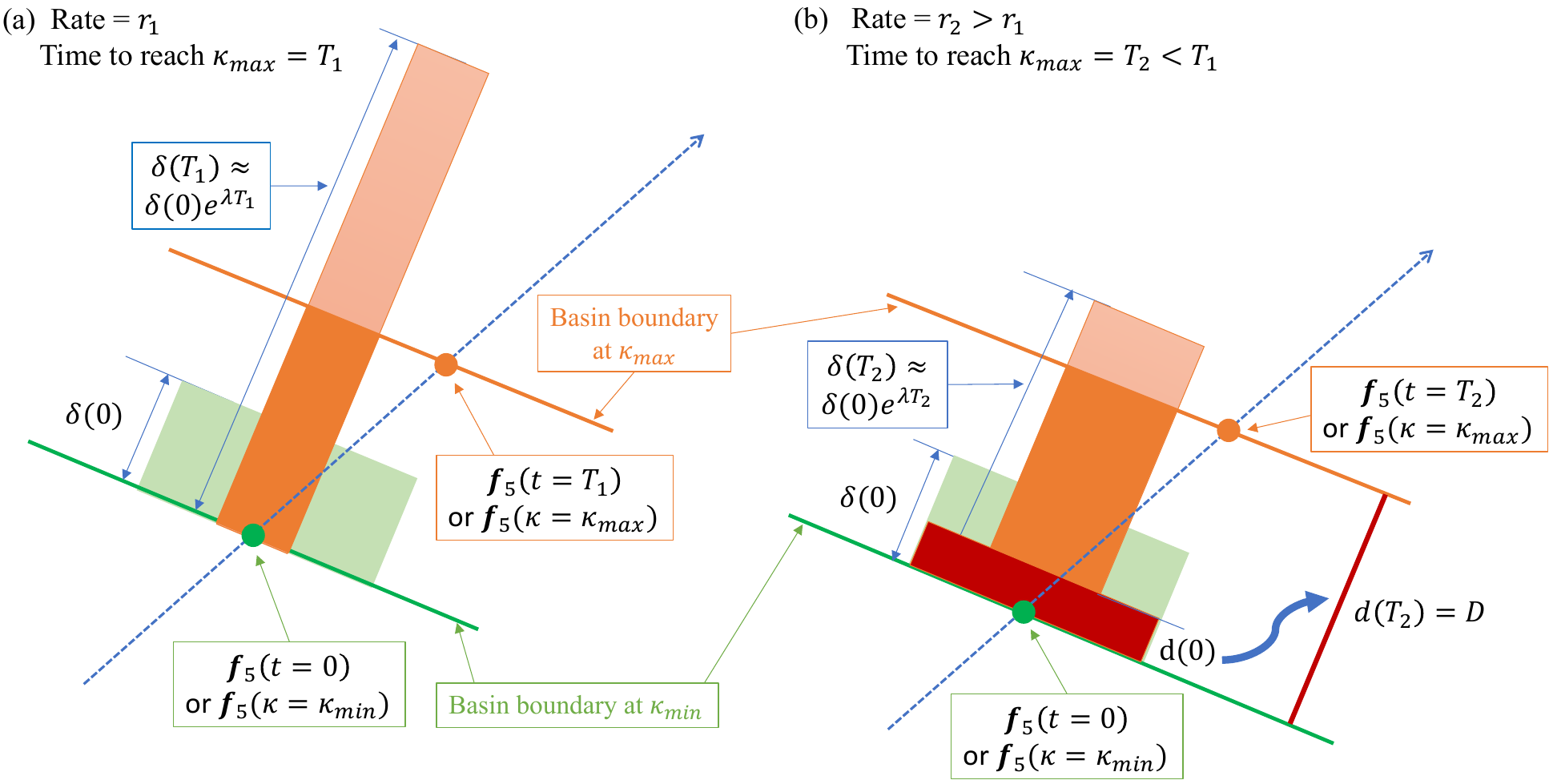}
\caption{Analysis of the dynamical mechanism responsible for R-tipping. The 
dynamics of the unstable fixed point $\mathbf{f}_5$ together with those of a 
rectangle region of the initial conditions in its neighborhood are illustrated 
for two values of the rate of parameter change: (a) $r = r_1$ and (b) 
$r = r_2 > r_1$. At $t=0$ ($\kappa = \kappa_{\rm min}$, a segment of the basin 
boundary in the vicinity of $\mathbf{f}_5$ is illustrated as the solid 
green line. As the rate of parameter change begins to increase, the basin 
boundary moves upward and eventually accumulates at that for 
$\kappa = \kappa_{\rm max}$ - the solid orange lines. For a small rate, the 
time required for the green boundary to reach the orange boundary is longer. 
All the initial conditions in the light green region above the green 
boundary belong to the basin of the survival attractor for 
$\kappa = \kappa_{\rm min}$, whose trajectories are determined by the local 
exponentially contracting and expanding dynamics of $\mathbf{f}_5$. The fate 
of these initial conditions is the result of a ``race'' to go above the orange 
basin boundary within the ``allowed'' time duration [$T_1$ in (a) and 
$T_2 < T_1$ in (b)]: those managing to go above will approach the survival 
attractor, but those that stay below will lead to extinction. Since $T_1 > T_2$,
the fraction of initial conditions that can go above in (a) is larger than that
in (b), leading to a higher probability of R-tipping for case (b). The precise 
fraction of those initial conditions is indicated by the red rectangular region
in (b), whose initial height stretches to the distance between the green and 
orange boundaries in the permissible time. This gives the probability of 
R-tipping as quantified by the scaling law (\ref{eq:scaling}). See text for 
more details.}
\label{fig:theory}
\end{figure*}

What will happen to the dynamics when the parameter $\kappa$ becomes time 
dependent? Without loss of generality, we focus on the interval of $\kappa$ as 
exemplified by the horizontal range of the dot-dashed ellipse in 
Figs.~\ref{fig:equi}(a) and \ref{fig:equi}(b), defined as
$[\kappa_{\rm min},\kappa_{\rm max}]$, in which there is bistability in the 
original high-dimensional network and in the 2D reduced model as well. Note 
that, for different high-dimensional empirical networks, the values of 
$\kappa_{\rm min}$ and $\kappa_{\rm max}$ in the 2D effective models are
different, as illustrated in the fifth column of Tab.~\ref{Tab:Net}. Because
of the coexistence of two stable fixed points, for every parameter value in the
range $[\kappa_{\rm min},\kappa_{\rm max}]$, there are two basins of 
attraction, as illustrated in Figs.~\ref{fig:equi}(c) and \ref{fig:equi}(d) 
for the 2D effective model of an empirical network ${\rm M}\_{\rm PL}\_036$ 
for $\kappa = 0.74$ and $\kappa = 0.88$, respectively, where the pink and gray 
regions correspond to the basins of the extinction and survival attractors,
respectively. The basin boundary is the stable manifold of the unstable 
fixed point $\mathbf{f}_5$. It can be seen that, as $\kappa$ increases, the 
basin of the extinction attractor increases, accompanied by a simultaneous 
decrease in the basin area of the survival attractor. This can be understood by 
comparing the positions of the equilibria in Fig.~\ref{fig:equi} where, as
$\kappa$ increases, the position of survival fixed point $\mathbf{f}_4$ moves 
towards lower plant and pollinator abundances, but the unstable fixed point 
$\mathbf{f}_5$ moves in the opposite direction: the direction of larger species
abundances. 

The boundary separating the basins of the extinction and survival fixed-point
attractors is the stable manifold of $\mathbf{f}_5$. As $r$ increases from 
zero, $\mathbf{f}_5$ moves in the direction of larger plant and pollinator
abundances, so must the basin boundary, as exemplified in Fig.~\ref{fig:bb} 
(the various dashed curves) for $r\in [0,1]$. Note that, since the decay 
parameter $\kappa$ increases from $\kappa_{\rm min}$ to $\kappa_{\rm max}$ 
at the linear rate $r$, we have $\kappa(r=0)=\kappa_{\rm min}$ and 
$\kappa(r=\infty)=\kappa_{\rm max}$. As $r$ increases, the basin boundaries 
accumulate at the one for $\kappa_{\rm max}$.

\section*{Theory}

Figure~\ref{fig:bb} provides a physical base for deriving the scaling law
(\ref{eq:scaling}). Consider the small neighborhood of the unstable fixed point 
$\mathbf{f}_5$, where the basin boundary is approximately straight, as shown
schematically in Fig.~\ref{fig:theory}. Consider two cases: one with rate $r_1$
of parameter increase and another with rate $r_2$, where $r_2 > r_1$, as shown
in Figs.~\ref{fig:theory}(a) and \ref{fig:theory}(b), respectively. For any 
given rate $r$, at the beginning and end of the parameter variation, we have 
$\kappa=\kappa_{\rm min}$ and $\kappa=\kappa_{\rm max}$, respectively, where 
the time it takes to complete this process is 
$T = (\kappa_{\rm max} - \kappa_{\rm min})/r$. Since $r_2 > r_1$, we have
$T_2 < T_1$. For both Figs.~\ref{fig:theory}(a) and \ref{fig:theory}(b), the
unstable fixed point $\mathbf{f}_5$ is marked by the filled green circles at
$t = 0$ and filled orange circles at the end of the parameter variation, 
and the blue dashed line with an arrow indicates the direction of change in 
the location of $\mathbf{f}_5$ in the phase space as the parameter varies with
time. Likewise, the solid green (orange) line segments through $\mathbf{f}_5$ 
denote the boundary separating the extinction basin from the survival basin of 
attraction at $t=0$ ($t=T_1$ or $T_2$). That is, before the parameter variation
is turned on for $\kappa=\kappa_{\rm min}$, the initial conditions below 
(above) the solid green lines belong to the basin of the extinction (survival) 
attractor. After the process of parameter variation ends so that 
$\kappa=\kappa_{\rm max}$, the initial conditions below (above) the solid 
orange lines belong to the basin of the extinction (survival) attractor. 
During the process of parameter variation, $\mathbf{f}_5$ moves from the 
position of the green circle to that of the orange circle, and its stable 
manifold (the basin boundary) moves accordingly. Now consider the initial 
conditions in the light shaded green area, which belong to the basin of the 
survival attractor for $\kappa=\kappa_{\rm min}$. Without parameter variation, 
as time goes, this green rectangular area will be stretched along the unstable 
direction of $\mathbf{f}_5$ exponentially according to its unstable eigenvalue
$\lambda$ and compressed exponentially in the stable direction, evolving into 
an orange rectangle that is long in the unstable direction. Since $T_1 > T_2$, 
the orange rectangle for $r = r_1$ is longer and thinner than that for 
$r = r_2$. 

The dynamical mechanism responsible for R-tipping can now be understood based
on the schematic illustration in Figs.~\ref{fig:theory}(a) and 
\ref{fig:theory}(b). In particular, because of the movement of $\mathbf{f}_5$ 
and the basin boundary as the parameter variation is turned on, the dark shaded
orange part of the long rectangle now belongs to the basin of the extinction 
attractor. The initial conditions in the original green rectangle which evolve 
into this dark shaded orange region are nothing but the initial conditions 
that switch their destinations from the survival to the extinction attractor
as the result of the time variation of the parameter. That is, these initial 
conditions will experience R-tipping, as indicated by the red rectangle inside
the green area in Fig.~\ref{fig:theory}(b). For any given rate $r$, the 
fraction of such initial conditions determines the R-tipping probability. 
Let $d(0)$ denote the fraction of R-tipping initial conditions
and let $D$ be the distance between the basin boundaries at the beginning and 
end of parameter variation along the unstable direction of $\mathbf{f}_5$. 
We have $d(T)=D=d(0)\exp{(\lambda T)}$. Our argument for $\Phi (r)\sim d(0)$
and the use of $T = (\kappa_{\rm max} - \kappa_{\rm min})/r$ lead to the 
scaling law (\ref{eq:scaling}).

\section*{Discussion}

{To summarize, nonlinear dynamical systems in nature such as ecosystems and 
climate systems on different scales are experiencing parameter changes due to 
increasing human activities, and it is of interest to understand how the 
``pace'' or rate of parameter change might lead to catastrophic consequences.
To this end, we studied high-dimensional mutualistic networks,
as motivated by the following general 
considerations. In ecosystems, mutualistic interactions, broadly defined as a 
close, interdependent, mutually beneficial relationship between two species, 
are one of the most fundamental interspecific relationships. Mutualistic networks 
contribute to biodiversity and ecosystem stability. As species within these 
networks rely on each other for essential services, such as pollination, seed 
dispersal, or nutrient exchange, they promote species coexistence and reduce 
competitive exclusion. This coexistence enhances the overall diversity of the 
ecosystem, making it more resilient to disturbances and less susceptible to the 
dominance of a few species. They can drive co-evolutionary processes between 
interacting species. As species interact over time, they may evolve in response 
to each other's adaptations, leading to reciprocal changes that strengthen the 
mutualistic relationship. Disruptions to these networks, such as the decline of 
pollinator populations, can have cascading effects on ecosystem functions and 
the survival of dependent species. By studying mutualistic interactions, 
conservationists can design more effective strategies to protect and restore 
these vital relationships and the ecosystems they support.}

This paper focuses on the phenomenon of rate-induced tipping or R-tipping, where
the rate of parameter change can cause the system to experience a tipping 
point from normal functioning to collapse. The main accomplishments are three. 
First, we went beyond the existing local approaches to R-tipping by taking a 
global approach of dynamical analysis based on consideration of basins of 
attraction of coexisting attractors. This allows us to introduce the 
probability of R-tipping with respect to initial conditions taken from the 
whole phase space. Second, most previous works on R-tipping analyzed 
low-dimensional toy models but our study focused on high-dimensional 
mutualistic networks constructed from empirical data. Third, we developed a 
geometric analysis and derived a scaling law governing the probability of 
R-tipping with respect to the rate of parameter change. The scaling law 
contains two parameters which, for two-dimensional systems, can be determined
theoretically from the dynamics. For high-dimensional systems, the two scaling
parameters can be determined through numerical fitting. For all {ten}
empirical networks studied, the scaling law agrees well with the results from direct 
numerical simulations. To our knowledge, no such quantitative law characterizing 
R-tipping has been uncovered previously. 

{The effective 2D systems in our study were previously derived~\cite{JHSLGHL:2018}, 
which capture the bipartite and mutualistic nature of the ecological interactions 
in the empirical, high-dimensional networks. It employs one collective variable to 
account for the dynamical behavior of the pollinators and another for the plants. 
The effective 2D models allow us to mathematically analyze the global R-tipping
dynamics, with results and the scaling law verified by direct simulations of
high-dimensional networks. The key reason that the dimension-reduced 2D models
can be effective lies, again, in the general setting of the ecological systems
under consideration: coexistence of a survival and an extinction state. While our
present work focused on mutualistic networks, we expect the global approach to
R-tipping and the scaling law to be generally applicable to ecological systems
in which large-scale extinction from a survival state is possible due to even a
small, nonzero rate of parameter change.}

{It is worth noting that the setting under which the scaling law of the probability 
of R-tipping with the rate of parameter change holds is the coexistence of two 
stable steady states in the phase space, one associated with survival and another 
corresponding to extinction. This setting is general for studying tipping, system 
collapse and massive extinction in ecological systems. Our theoretical analysis 
leading to the scaling law requires minimal conditions: two coexisting basins of 
attraction separated by a basin boundary. The scaling law is not the result of some 
specific parameterization of the mutualistic systems but is a generic feature in 
systems with two coexisting states. Insofar as the system can potentially undergo a 
transition from the survival to the extinction state, we expect our R-tipping 
scaling law to hold. In a broad context, coexisting stable steady states or 
attractors are ubiquitous in nonlinear physical and biological systems.} 

The scaling law stipulates that, as the rate increases from zero, the R-tipping 
probability increases rapidly first and then saturates. This has a striking and 
potentially devastating consequence: in order to reduce the probability of 
R-tipping, the parameter change must be slowed down to such an extent that its 
rate of change is practically zero. This has serious implications. For example, 
to avoid climate-change induced species extinction, it would be necessary to ensure 
that no parameters change with time, and this may pose an extremely significant
challenge in our efforts to protect and preserve the natural environment.

\acknow{This work was supported by the Office of Naval Research under Grant No.~N00014-21-1-2323. Y. Do was supported by the National Research Foundation of Korea (NRF) grant funded by the Korean government (MSIP) (Nos.~NRF-2022R1A5A1033624 and 2022R1A2C3011711).} 

\showacknow{}

\end{document}